\documentclass[sigconf]{acmart}
\AtBeginDocument{%
  }

\copyrightyear{2026}
\acmYear{2026}
\setcopyright{cc}
\setcctype{by}
\acmConference[CHI '26]{Proceedings of the 2026 CHI Conference on Human Factors in Computing Systems}{April 13--17, 2026}{Barcelona, Spain}
\acmBooktitle{Proceedings of the 2026 CHI Conference on Human Factors in Computing Systems (CHI '26), April 13--17, 2026, Barcelona, Spain}
\acmPrice{}
\acmDOI{10.1145/3772318.3790411}
\acmISBN{979-8-4007-2278-3/2026/04}

\usepackage{stfloats}
\usepackage{tabularx} 
\usepackage{enumitem}   
\usepackage{marginnote} 
\usepackage{multirow}
\usepackage{graphicx}
\usepackage{caption}      
\usepackage{framed}
\usepackage{tabularx,booktabs,ragged2e,makecell}
\newcolumntype{L}{>{\RaggedRight\arraybackslash}X}
\newcolumntype{C}{>{\centering\arraybackslash}p{1.7cm}}

\begin{document}
\author{Kaori Ikematsu}
\orcid{0000-0002-7017-6744}
\affiliation{%
  \institution{LY Corporation}
  \city{Tokyo}
  \country{Japan}}
\authornotemark[1]
\email{k-ikematsu@acm.org}

\author{Kunihiro Kato}
\orcid{0000-0002-0117-8981}
\affiliation{%
  \institution{Tokyo University of Technology}
  \city{Tokyo}
  \country{Japan}}
\authornote{Both authors contributed equally to this work.}
\email{kkunihir@acm.org}

\title{DuoTouch: Passive Two-Footprint Attachments Using Binary Sequences to Extend Touch Interaction}
\renewcommand{\shortauthors}{Ikematsu and Kato}

\begin{abstract}
DuoTouch is a passive attachment for capacitive touch panels that adds tangible input while minimizing content occlusion and loss of input area. 
It uses two contact footprints and two traces to encode motion as binary sequences and runs on unmodified devices through standard touch APIs. 
We present two configurations with paired decoders: an aligned configuration that maps fixed-length codes to discrete commands and a phase-shifted configuration that estimates direction and distance from relative timing. 
To characterize the system's reliability, we derive a sampling-limited bound that links actuation speed, internal trace width, and device touch sampling rate. 
Through technical evaluations on a smartphone and a touchpad, we report performance metrics that describe the relationship between these parameters and decoding accuracy. 
Finally, we demonstrate the versatility of DuoTouch by embedding the mechanism into various form factors, including a hand strap, a phone ring holder, and touchpad add-ons. 
\end{abstract}



\begin{teaserfigure}
\includegraphics[width=\textwidth]{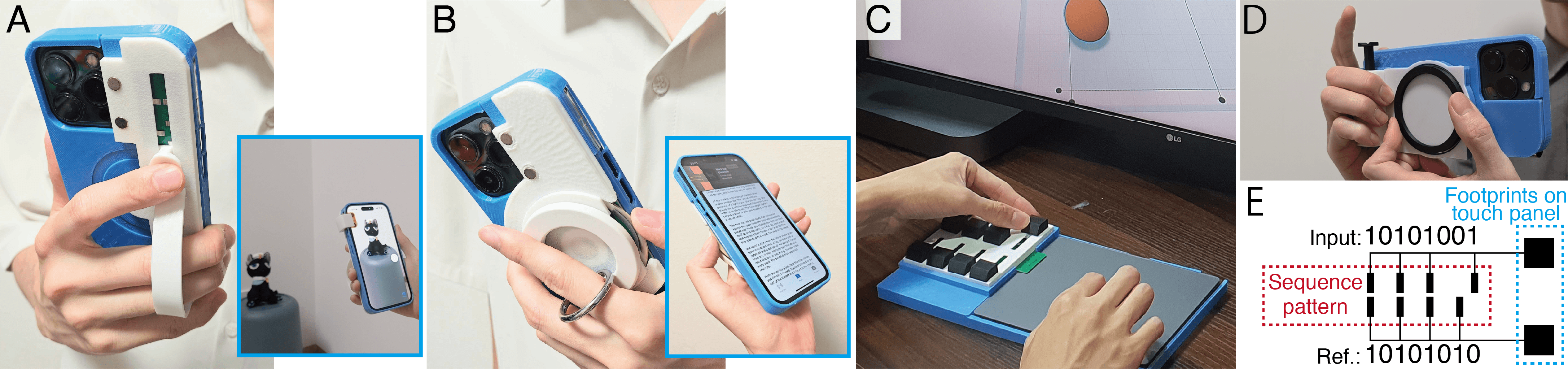}
\caption{
(A--D) DuoTouch applications and (E) a sample of a sequence pattern that encodes motion into binary sequences.
}
\Description{(A--D) Four photos showing DuoTouch prototypes integrated into a smartphone hand strap, a phone ring holder, and touchpad add-ons. (E) A diagram of a sequence pattern consisting of two contact footprints and traces designed to encode motion into binary sequences.}
  \label{fig:teaser}
\end{teaserfigure}

\maketitle
\section{Introduction}
Passive attachments using conductive materials have long been explored in both human-computer interaction (HCI) research and the industry at large to extend touch interactions~\cite{ExtensionSticker,ShiftTouch,KirigamiKeyboard,PaperTouch,Clip-onGadgets,CAPath,BackTrack,CapWidgets,PUCs,Anamorphicons,SurfaceConstellations,PERCs,CapCodes,CapStones,Transporters,SheetKey}. While touch panels are primarily designed for finger input, they can also detect grounded conductive objects (e.g., metal or conductive rubber)~\cite{Ohmic-Sticker,Ohmic-Touch}. 
This principle has been applied to commercial products such as keyboard overlays\footnote{Galaxy Note5 Keyboard Cover (Samsung) | \url{https://www.samsung.com/us/mobile/mobile-accessories/phones/galaxy-note5-keyboard-cover-black-sapphire-ej-cn920ubegus/}} and trigger attachments for mobile gaming\footnote{PuK Triggers (PuK Gaming) | \url{https://www.pukgaming.com/products/puk-triggers}}. 
In addition, interfaces have been proposed that enable gesture-based input, including scrolling, pinching, and pointing, by physically extending the touch-sensitive area~\cite{CAPath,BackTrack,ExtensionSticker}, or by incorporating input structures such as buttons, sliders, or dials~\cite{PaperTouch,Clip-onGadgets,CapWidgets,CapStones}. 
These interfaces operate passively and incur low maintenance, monetary, and implementation costs, making them promising for retrofitting existing devices and wide deployment. 
However, supporting rich interactions requires multiple on-screen footprints (one per input~\cite{Clip-onGadgets,PaperTouch}), which increases visual occlusion and makes it difficult to see the screen and perform precise finger input~\cite{ShiftTouch}. Although some interfaces~\cite{ShiftTouch,KirigamiKeyboard} reduce the number of required footprints by recognizing unique combinations of multiple footprints (i.e., enabling more inputs than the number of footprints alone), a considerable number of footprints are still necessary to support a wide range of inputs. Furthermore, the complexity of wiring between footprints and user touch points can increase implementation costs~\cite{ShiftTouch}.

To address these issues, we introduce DuoTouch, a passive attachment that enables recognition of  diverse user actions, such as sliding, rotating, and pushing, using only two conductive on-screen footprints and their two traces (Fig.~\ref{fig:teaser}). This simple structure supports a wide range of input actions while minimizing visual occlusion and wiring complexity. The attachment contains two conductive traces, collectively called a {\it sequence pattern} (Fig.~\ref{fig:teaser}E), that run from a pair of on-screen footprints toward the user-actuated component. 
User actions along these traces produce binary touch sequences (temporal patterns of touches and releases on the footprints), which can represent either discrete codes or continuous movement patterns. By analyzing the detected sequence, DuoTouch can either identify the interaction type to trigger discrete commands or measure the direction and distance of user motion for continuous control, depending on the trace configuration. Because it relies solely on standard capacitive touch sensing, DuoTouch works across smartphones, tablets, touchpads, and other form factors on mainstream operating systems, which makes it a practical interface but requires addressing three key design challenges: 
\begin{enumerate}[label=C\arabic*., leftmargin=14pt]
\item \textbf{Encoding and decoding with only two footprints.} How can the sequence pattern and the decoder be organized so that both discrete command selection and continuous control are feasible on commodity touch controllers without firmware changes? 
\item \textbf{Operating envelope on commodity panels.} What geometric and sampling conditions bound reliable operation across devices and electrode sizes?
\item \textbf{From principle to practice.} How can these conditions be turned into repeatable design choices and a tool that produces valid patterns across form factors?
\end{enumerate}

We address C1 by introducing two complementary decoding strategies on a shared two-trace architecture. The aligned configuration maps motion to fixed-length binary sequences for discrete commands. The phase-shifted configuration estimates direction and distance from the relative timing between the two traces for continuous control. Both strategies work with standard touch APIs and do not require kernel or firmware changes. For C2, we derive a sampling-limited bound that governs when motion can be resolved and show that it scales with electrode width and sensing rate. We validate these results with an empirical evaluation on a smartphone and a touchpad across multiple widths and speeds. For C3, we distill the findings into device-independent design guidelines expressed in a normalized speed measure, and we implement a design-support tool that generates patterns satisfying the validated criteria and exports them in CAD-ready formats. We also implement application examples on both device classes to illustrate how DuoTouch scales from discrete commands to continuous control, and we gather preliminary user feedback to inform design choices. 
Ikematsu et al.~\cite{DuoTouch_uist} introduced the initial concept of DuoTouch without reporting accuracy validation, user study, or a design-support tool; building on this concept, we address these gaps. 
Our contributions are summarized as follows:
\begin{itemize}[noitemsep,leftmargin=10pt]
\item \textbf{DuoTouch}, a passive two-footprint attachment that encodes motion as binary touch sequences on unmodified capacitive touch panels.
\item \textbf{Two configurations with paired decoders}: an \emph{aligned configuration} with a \emph{code classifier} for discrete commands, and a \emph{phase-shifted configuration} with a \emph{geometric estimator} for continuous control.
\item \textbf{A technical evaluation on a smartphone and a touchpad} across four electrode widths and a range of speeds, revealing accuracy trends and boundary conditions. 
\item \textbf{An analytical characterization} that links input speed, electrode width, and sampling rate into a sampling-limited bound that explains the operating envelope. 
\item \textbf{Design guidelines and a design-support tool} that operationalize the findings with a normalized speed measure and export verified patterns in CAD-ready vector formats.
\end{itemize}

\section{Related Work}

\subsection{Extending Touch Panel Interactions with Passive Attachments}

Conventional methods for extending touch panel interactions face challenges with screen occlusion and wiring complexity. 
Table~\ref{tab:comparison} summarizes representative methods for each interaction type and their trade-offs. 
For example, interfaces that require one-to-one connections between input areas and the touch panel (e.g., Clip-on gadgets~\cite{Clip-onGadgets}, PaperTouch~\cite{PaperTouch}, and \cite{CapWidgets,CapCodes,CapStones})
increase the number of physical attachments as more inputs are added. This increases the on-screen footprint and can obstruct the display and affect usability~\cite{ShiftTouch}. 
Kirigami Keyboard~\cite{KirigamiKeyboard} routes each key to shared footprints and identifies it from the active set, 
while BackTrack~\cite{BackTrack} maps each touch on the back-of-device touchpad to combinations of front-screen electrodes. 
While these schemes reduce footprint size relative to strict one-to-one mappings, they still require many footprints to support large key sets~\cite{KirigamiKeyboard} or high-resolution continuous tracking~\cite{BackTrack}. 
ShiftTouch~\cite{ShiftTouch} uses multiple linear electrodes (a strip array) as on-screen footprints to sense lateral shifts in touch position. Shared electrodes reduce area but still leave many small footprints and increase the wiring complexity. 
In contrast, DuoTouch uses two on-screen footprints regardless of the number of inputs, and each footprint forms a single connected sensing area, simplifying the wiring.  
Additionally, ExtensionSticker~\cite{ExtensionSticker} uses strip arrays of footprints for 1D scrolling, and CAPath~\cite{CAPath} uses grid footprints for 2D pointing. 
While these interfaces function via direct touch coordinate transmission, DuoTouch relies on a decoder that maps simple binary sequences from its two fixed footprints. 
This encoding supports many discrete commands determined by the code length and also supports continuous input.

\begin{table*}[h]
\Small
\setlength{\tabcolsep}{5pt}
\renewcommand{\arraystretch}{1.05}
\caption{
Comparison of screen-attached passive interfaces. 
\textcolor{green!80!black}{Green} marks favorable properties: 
\emph{Wiring scaling} indicates the asymptotic growth of wiring as a function of the on-screen footprint count $T_{\mathrm{fp}}$; 
\emph{Mapping} indicates whether a decoder is required (No means direct touch coordinate transmission);
\emph{Public API} indicates compatibility with off-the-shelf OS APIs without private or raw data access; 
\emph{Authoring tool} indicates the availability of a standalone software generator or interface for creating the interface assets (to distinguish from systems providing only methodological guidelines); 
\emph{Manual} indicates whether hand assembly is required. 
}
\Description{
Comparison of screen-attached passive interfaces. The table evaluates technical properties such as wiring scaling, Public API compatibility, and the availability of an authoring tool. 
}
\label{tab:comparison}
\begin{tabularx}{\textwidth}{@{}%
  >{\raggedright\arraybackslash}p{0.13\textwidth}%
  >{\raggedright\arraybackslash}p{0.18\textwidth}%
  >{\raggedright\arraybackslash}p{0.07\textwidth}%
  >{\raggedright\arraybackslash}p{0.06\textwidth}%
  >{\raggedright\arraybackslash}p{0.06\textwidth}
  >{\raggedright\arraybackslash}p{0.04\textwidth}%
  >{\raggedright\arraybackslash}p{0.06\textwidth}%
  >{\raggedright\arraybackslash}p{0.05\textwidth}%
  >{\raggedright\arraybackslash}X@{}}
\toprule
\textbf{Approach} & \textbf{Interaction type} & \textbf{Footprints} & \textbf{Wiring scaling} & \textbf{Mapping} & \textbf{Public API} & \textbf{Authoring tool} & \textbf{Manual} & \textbf{Fabrication} \\
\midrule
\textbf{DuoTouch (ours)} & Continuous (1D) \& Discrete & \textbf{\textcolor{green!80!black}{2 (fixed)}} &
\textbf{\textcolor{green!80!black}{O(1)  
}} 
& Yes &
\textbf{\textcolor{green!80!black}{Yes}} & \textbf{\textcolor{green!80!black}{Yes}} & Yes & PCBs; 3D printed parts \\
Clip-on Gadgets~\cite{Clip-onGadgets} & Discrete & 1/input & \(O(T_{\mathrm{fp}})\) & Yes &
\textbf{\textcolor{green!80!black}{Yes}} & No & Yes & Conductive tacks; rubber pads\\
PaperTouch~\cite{PaperTouch} & Discrete & 1/input & \(O(T_{\mathrm{fp}})\) & Yes &
\textbf{\textcolor{green!80!black}{Yes}} & No & Yes & Hand drawn conductive ink\\
Kirigami Keyboard~\cite{KirigamiKeyboard} & Discrete & Combined & \(O(T_{\mathrm{fp}})\) & Yes &
\textbf{\textcolor{green!80!black}{Yes}} & No & Yes & Printed conductive ink on film \\
ShiftTouch~\cite{ShiftTouch} & Discrete & Strip array & \(O(T_{\mathrm{fp}})\) & Yes &
\textbf{\textcolor{green!80!black}{Yes}} & No & Yes & Printed conductive ink on film \\
ExtensionSticker~\cite{ExtensionSticker} & Continuous (1D) & Strip array & \(O(T_{\mathrm{fp}})\) &
\textbf{\textcolor{green!80!black}{No}} &
\textbf{\textcolor{green!80!black}{Yes}} & No & \textbf{\textcolor{green!80!black}{No}} & Printed conductive ink on film \\
CAPath~\cite{CAPath} & Continuous (2D) & Grid array & \(O(T_{\mathrm{fp}})\) &
\textbf{\textcolor{green!80!black}{No}} &
\textbf{\textcolor{green!80!black}{Yes}} & \textbf{\textcolor{green!80!black}{Yes}} & \textbf{\textcolor{green!80!black}{No}} & 3D printed parts \\
BackTrack~\cite{BackTrack} & Continuous (2D) & Combined & \(O(T_{\mathrm{fp}})\) & Yes &
\textbf{\textcolor{green!80!black}{Yes}} & No & Yes & ITO film; PCBs; 3D printed parts \\
FlexTouch~\cite{FlexTouch} & Continuous (1D/2D) \& Discrete & Strip array & \(O(T_{\mathrm{fp}})\) & Yes &
No & No & Yes & Conductive paints, inks, or tapes \\
Ohmic-Sticker (TrackPoint-type)~\cite{Ohmic-Sticker} & Continuous (2D+) & 4 (fixed) & \textbf{{\textcolor{green!80!black}{O(1)}}} & Yes &
No & No & Yes & PCBs; conductive film \\
\bottomrule
\end{tabularx}
\end{table*}

Another approach to reducing the number or area of on-screen footprints is to read raw capacitive data from the touch controller~\cite{Ohmic-Touch,Ohmic-Sticker,Flip-FlopSticker,LightTouch,FlexTouch,Flexibles,Itsy-Bits}. 
Ohmic-Sticker~\cite{Ohmic-Sticker} and Flexibles~\cite{Flexibles} modulate impedance to ground so that a single footprint yields a continuous signal for press-like force. 
Ohmic-Touch~\cite{Ohmic-Touch} detects various user actions, including bending and rotation, and FlexTouch~\cite{FlexTouch} senses proximity and long-range presence. 
While these interfaces enable analog sensing on and around the touch panel, they typically rely on non-public capacitive images or modalities~\cite{Ohmic-Touch,Ohmic-Sticker} and often require driver or firmware modifications~\cite{Flexibles,FlexTouch}. 
DuoTouch instead targets unmodified touch panels and decodes from standard touch APIs.

For these interfaces, rapid prototyping and flexibility for custom designs are also essential. 
Prior work has demonstrated low-cost prototyping with inkjet printing~\cite{ExtensionSticker,KirigamiKeyboard,ShiftTouch} and single-step fabrication of custom conductive paths using multi-material 3D printers~\cite{Itsy-Bits,CAPath}. 
While DuoTouch requires manual assembly of PCBs into a 3D-printed housing, it provides a design pipeline with a sequence-pattern design-support tool and a library of STL files for 3D-printed parts.

\subsection{Binary Sequences in Interaction Techniques}
Binary sequences encode information as ordered patterns of discrete events. 
They offer two advantages for interaction design. 
First, since only the timing and order of on and off events matter, raw signals can be reduced to binary transitions, which reduces dependence on precise amplitude measurements and improves robustness to noise and environmental changes. 
Second, patterned sequences can express many commands with few events, and their discrete structure maps cleanly to finite-state machines, making recognition logic straightforward. 
These benefits have motivated binary encodings in diverse sensing modalities, including acoustic signals from notched surfaces~\cite{AcousticBarcodes}, timed accelerometer peaks from knuckle knocks~\cite{Yourknock}, micro-scale structures that modulate reflected light~\cite{StructCode,AnisoTag,AirCode,InfoPrint}, and bursts of infrared reflection from a finger-worn ring~\cite{iRing}. 
We focus on capacitive touch sensing because commercially available controllers already scan the panel during normal use, enabling decoding without hardware or firmware changes and with lightweight software processing. 

Several closely related systems treat touch contact as binary data and encode temporal sequences as codes, such as FlashTouch~\cite{FlashTouch} for data transmission and TUIC~\cite{TUIC} for identifying tangibles. These systems use microcontroller-driven switching circuits to synthesize touch events, whereas DuoTouch also uses binary sequences but relies only on passive structures. 
Complementary to sequence-based methods, MonoTouch~\cite{MonoTouch} distinguishes touch gestures from the analog time series of a dedicated single-electrode capacitive sensor, whereas DuoTouch represents interactions as binary sequences decoded from touch events on the commodity capacitive touch panel.

\subsection{Extending Interaction Regions with On-Device Sensing and Hardware Attachments}
Research on extended input spaces explores off-screen, around-device~\cite{BeyondTouch,FingerIO,Acoustic+Pose,GlassHands}, and back-of-device~\cite{TapNet,BackXPress} interaction using various on-device sensors and hardware augmentations. 
On-device sensing examples include BeyondTouch~\cite{BeyondTouch}, which enables off-screen taps and slides using the accelerometer, gyroscope, and microphones. 
TapNet~\cite{TapNet} also enables off-screen taps but relies solely on inertial signals. 
Other approaches use active acoustic sensing to probe the near-field via the speaker-microphone path~\cite{FingerIO,SmartGrip,Acoustic+Pose,Screen-IntegratedSpeakers} and vision-based approaches that utilize the user-facing camera for detecting device grasp~\cite{ReflecTouch,HandyTrak} and near-device interactions~\cite{GlassHands,ReflecTrace}. 
Hardware-augmented systems offer another approach by coupling physical mechanisms to the device. 
Some provide tangible controls, such as Acoustruments~\cite{Acoustruments}, which routes passive controls via the audio path, and MagPie~\cite{MagPie}, which provides back-of-device controls that generate distinctive signals. 
Others add new sensing modalities, such as BackXPress~\cite{BackXPress}, which senses back-of-device finger pressure to complement front touch. 
While these approaches are promising, many rely on sensors that are not continuously active by default, thereby increasing the energy consumption and runtime costs and in some cases raising privacy concerns. 
They are also susceptible to environmental noise (e.g., ambient sound or lighting) or incidental device motion. In contrast, our approach operates entirely within the capacitive channel, which is inherently more robust to environmental factors and allows for a simple, calibration-free algorithm.

\section{Design Space}
\label{sec:design-space}
\begin{figure*}[b]
  \includegraphics[width=\textwidth]{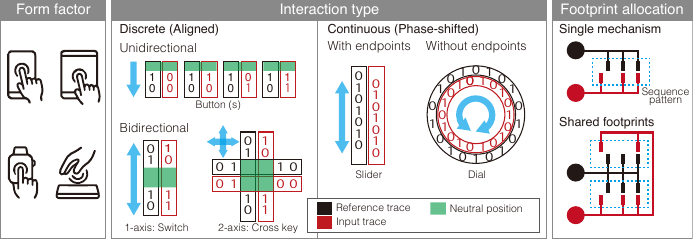}
  \caption{
  DuoTouch design space by form factor, interaction type, and footprint allocation; green marks the neutral position.  }
\Description{Three columns illustrating the DuoTouch design space: Form Factor (phone, tablet, watch, and touchpad icons); Interaction Type, divided into Discrete (aligned) for buttons, switches, and cross keys, and Continuous (phase-shifted) for sliders and dials; and Footprint Allocation, showing wiring for single mechanisms and shared footprints that route multiple mechanisms to the same two contacts.}
  \label{fig:PrimitiveInteractions}
\end{figure*}

DuoTouch is a passive attachment for commodity touch panels that extends touch interaction using two contact footprints via standard touch APIs 
across smartphones, tablets, smartwatches, and touchpads. 
We define a design space that organizes the design primitives of DuoTouch: architecture and trace geometry, configuration, and footprint allocation (Fig.~\ref{fig:PrimitiveInteractions}). 
The box below establishes the notation used throughout the paper. 
\begin{framed}
\noindent \textbf{Notation} \\
\footnotesize
\begin{tabularx}{\linewidth}{lX}
$w$ & electrode width \ [mm] \\
$L_{\mathrm{seq}}$ & full-cycle spatial pitch of the reference pattern along the motion path \ [mm] \\
$f_{\mathrm{s}}$ & sampling rate (touch sensing frame rate) \ [Hz] \\
$T_s$ & sampling period, $T_s = 1/f_{\mathrm{s}}$ \ [s] \\
$v$ & actuation speed along the sequence pattern \ [mm/s] \\
$s$ & normalized speed, $s=\dfrac{v}{w f_{\mathrm{s}}}$ \ (dimensionless) \\
$\Delta d$ & minimum detectable movement per transition (phase-shifted only); 
$\ \Delta d=\dfrac{L_{\mathrm{seq}}}{2}$ in interleaved geometry 
\end{tabularx}
\end{framed}
Design parameters are chosen later by the workflow: namely, the width $w$ and spacing set the reference pitch $L_{\mathrm{seq}}$ and the motion resolution $\Delta d$; in the aligned configuration, the code length $L$ sets command capacity; and in the phase-shifted configuration, the target travel sets the required trace length.

\begin{figure*}[bht]
  \includegraphics[width=0.80\textwidth]{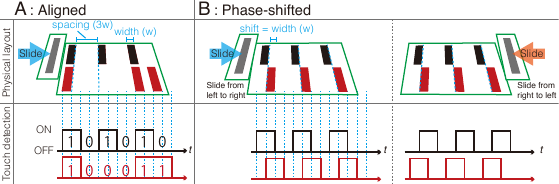}
\caption{
Examples of six-bit sequence patterns: (A) aligned and (B) phase-shifted (black = reference, red = input).  }
\Description{Six sequence patterns illustrating aligned and phase-shifted configurations. Each consists of a diagram showing the physical arrangement of reference and input traces, sliding directions, and the resulting binary on-off time series generated during interaction.}
    \label{fig:evaluation_pattern}
\end{figure*}

\paragraph{\textbf{Basic Architecture. }}
We first define the basic architecture of DuoTouch. 
It is composed of two classes of parts: PCBs (a main board for the contact footprints and the sequence pattern, and a small board for the user-actuated component) and 3D-printed parts for the housing and the user-actuated component. 
On the main PCB, two footprints are positioned over the active area of a commodity capacitive touch panel (Fig.~\ref{fig:teaser}E).  
From each footprint, a narrow conductive trace extends along the motion path of the user-actuated component. 
The two traces form a \emph{sequence pattern} composed of alternating conductive and insulating segments. 
A user-actuated component, such as a knob or button, contains an embedded conductive portion and moves along the traces during operation. 
When this conductive portion contacts a segment of a trace, it forms an electrical path from the touch panel through the trace, the conductive portion, and the user's body, allowing the touch controller to detect a touch at the corresponding footprint. 
Each footprint is reported by the touch controller as a standard touch point. 
Both traces use the same segment pitch to maintain a consistent temporal resolution. 
Depending on the intended interaction, the two traces can be placed in either a spatially aligned or a phase-shifted layout: 
\begin{itemize}[leftmargin=1.2em]
    \item \emph{Reference trace}:     
    A periodic sequence of alternating conductive and insulating segments with fixed pitch. 
    This pattern generates a binary timing signal (e.g., \colorbox{gray!20}{1010...}) as the user-actuated component moves. This timing information serves as the sampling reference for decoding.
    \item \emph{Input trace}: 
    Configured to encode either a fixed binary code or a phase-relative signal.  In aligned configurations (Fig.~\ref{fig:evaluation_pattern}A), the pattern encodes a discrete command; in phase-shifted configurations (Fig.~\ref{fig:evaluation_pattern}B), the relative timing between input and reference events is utilized to detect motion direction and distance. 
\end{itemize}

\paragraph{\textbf{Interaction Modes by Configuration. }}
We define the interaction modes realized by the two configurations. 
Table~\ref{tab:example-mappings} summarizes the mappings from interaction types to configurations, with the corresponding control and pattern assignment.
\begin{itemize}[leftmargin=1.2em]
  \item \emph{Aligned configuration (discrete)}:  
  This configuration follows a synchronous serial timing model, where the reference trace defines bit intervals, the input trace carries data, and the decoder latches one bit per interval~\cite{TheArtofElectronics,HandbookOfModernSensors}. 
  It supports unidirectional codes for simple actions (e.g., button presses) and bidirectional codes that distinguish opposing directions (e.g., toggle switches, cross keys). 
\item \emph{Phase-shifted configuration (continuous)}: This configuration follows a quadrature timing model, where the two traces form a \(90^{\circ}\) quadrature pair, direction is given by the lead or lag relation, and distance is obtained by counting transitions~\cite{TheArtofElectronics,HandbookOfModernSensors}.
It supports bounded sliders with endpoints and unbounded rotary dials without endpoints.
\end{itemize}

\begin{table*}[bht]
\centering
\Small
\caption{Mapping interaction types to controls, configurations, and pattern assignments.}
\Description{A table mapping five interaction types to their respective controls and configurations. Discrete interactions (one-shot, toggle, and directional) use the aligned configuration with buttons, switches, or cross keys. Continuous interactions (scalar and cyclic) use the phase-shifted configuration with sliders or dials.}
\label{tab:example-mappings}
\begin{tabular}{p{0.15\linewidth} p{0.3\linewidth} p{0.10\linewidth} p{0.12\linewidth} p{0.21\linewidth}}
\toprule
\textbf{Interaction type} & \textbf{Representative operations} & \textbf{Control} & \textbf{Configuration} & \textbf{Pattern assignment} \\
\midrule
One-shot action & Trigger shortcut; undo; capture screenshot & Button & Aligned & Unidirectional code \\
Two-state toggle & Toggle modes; play/pause; mute/unmute & Switch & Aligned & Reversible pair $\{p,\mathrm{rev}(p)\}$ \\
Directional navigation & Move focus; move caret; switch tabs & Cross keys & Aligned & Small codebook by direction \\
Scalar adjustment & Scroll; zoom; adjust volume; adjust brightness & Slider or dial & Phase-shifted & Bounded or unbounded travel \\
Cyclic adjustment & Rotate; adjust hue  & Dial & Phase-shifted & Unbounded travel \\
\bottomrule
\end{tabular}
\end{table*}

\paragraph{\textbf{Footprint allocation.}}
Multiple mechanisms can share the same two contact footprints. The decoder assumes only one mechanism is actuated at a time. This enables multiple interactions with minimal footprint and wiring overhead.

\section{Signal Generation and Decoding}
\label{sec:signal-generation-and-decoding}
As the user-actuated component moves, the touch panel reports \colorbox{gray!20}{1} when the conductive portion overlaps a conductive segment and \colorbox{gray!20}{0} when it crosses an insulating gap. 
The decoder interprets these signals using one of the following methods, depending on the trace alignment.

\subsection{Aligned decoding}
The decoder first divides the timeline into successive \emph{reference-bit intervals}; Using the notation in Sec.~\ref{sec:design-space}, let $t^{\mathrm{ref}}_{i}$ denote the start time of the $i$-th reference-bit interval. 
We write $T_{\mathrm{s}}{=}1/f_{\mathrm{s}}$
for the sampling period and use ``frame'' to mean a sampling frame. 
Let \(\mathcal{F}_i\) be the set of frames whose frame intervals overlap the \(i\)-th reference-bit interval, and let \(N_i = |\mathcal{F}_i|\).
Let \(x_n\in\{0,1\}\) be the observed touch state in frame \(n\), where the frame interval overlaps with the reference-bit interval.
For each interval the input bit \(b_i\) is decided by a plain majority over overlapping frames:
{\Small
\begin{equation}
b_i =
\begin{cases}
1, & \text{if } \sum_{n\in\mathcal{F}_i} x_n > \frac{1}{2} N_i,\\
0, & \text{if } \sum_{n\in\mathcal{F}_i} x_n < \frac{1}{2} N_i.
\end{cases}
\end{equation}
}

Note that if a tie occurs (\(\sum_{n\in\mathcal{F}_i} x_n = \tfrac{1}{2} N_i\)), actual frame timestamps vary around \(T_{\mathrm{s}}\) even when the nominal frame rate is fixed. Let \(\delta_n\) be the overlap duration between frame \(n\) and the interval. 
The tie is resolved as:
{\Small
\begin{equation}
b_i =
\begin{cases}
1, & \text{if } \sum_{n\in\mathcal{F}_i} x_n \,\delta_n > \tfrac{1}{2} \sum_{n\in\mathcal{F}_i} \delta_n\\
0, & \text{if } \sum_{n\in\mathcal{F}_i} x_n \,\delta_n < \tfrac{1}{2} \sum_{n\in\mathcal{F}_i} \delta_n
\end{cases}
\end{equation}
}
When equality holds after duration‑weighted tie‑breaking, the bit is marked as ambiguous. Repeating this computation for \(i=0,\ldots,L-1\) yields an \(L\)-position pattern that may include ambiguous positions. 
We then match this pattern to the predefined command codes of length \(L\), treating ambiguous positions as wildcards. 
A command is issued only when exactly one code matches.

\paragraph{\textbf{Assignable codes.}}
For a binary sequence of length \(L \ge 2\) there are \(2^{L}\) possible bit patterns. 
When bidirectional input is required, the decoder must discriminate a pattern from its reverse because a leftward swipe produces a time-reversed bit sequence of a rightward swipe. 
A pattern identical to its reverse, that is, a \emph{palindromic sequence} such as \colorbox{gray!20}{0110} produces the same bit pattern for leftward and rightward swipes and therefore cannot distinguish swipe direction.

\subsection{Phase-shifted decoding}
For each \emph{reference-bit interval} \([t^{\mathrm{ref}}_{i,\mathrm{start}},\, t^{\mathrm{ref}}_{i,\mathrm{end}})\), we detect input transitions of the same type as the corresponding reference edge.
Let \(t^{\mathrm{input}}_{i,\mathrm{start}}\) and \(t^{\mathrm{input}}_{i,\mathrm{end}}\) be the timestamps of the first such transitions that occur within, or immediately after, the interval.
Define the offsets
{\Small
\begin{equation}
\Delta t_{i,\mathrm{start}} = t^{\mathrm{input}}_{i,\mathrm{start}} - t^{\mathrm{ref}}_{i,\mathrm{start}}, \quad
\Delta t_{i,\mathrm{end}}   = t^{\mathrm{input}}_{i,\mathrm{end}}   - t^{\mathrm{ref}}_{i,\mathrm{end}}
\end{equation}
}

We call an offset \emph{neutral} when the input and reference transitions fall into the same sampling frame, that is, when the corresponding quantized timestamps coincide so that the offset equals zero. 
Concretely, the condition is $\Delta t_{i,\mathrm{start}}=0$ for the start edge and $\Delta t_{i,\mathrm{end}}=0$ for the end edge.
Otherwise that offset is signed with $\mathrm{sgn}(\Delta t_{i,\mathrm{start}})\in\{-1,+1\}$ or $\mathrm{sgn}(\Delta t_{i,\mathrm{end}})\in\{-1,+1\}$, respectively.
If no input transition of the required type is found within the interval or immediately after, the offset is undefined.

\paragraph{\textbf{Direction decision.}}
By convention, a negative offset means the input leads the reference and a positive offset means it lags.
Direction is identified when both offsets are signed with the same sign, and may also be identified when one offset is signed and the other is neutral or undefined. If the signs differ, the direction is not identified.

\paragraph{\textbf{Transition timing and attribution.}}
We detect transitions in the framewise binary stream from changes \(0\!\to\!1\) and \(1\!\to\!0\) between consecutive frames; each of these changes yields at most one transition event. We set the transition timestamp to the start time of the first frame after the state change. 
If a transition timestamp exactly matches a reference boundary, we attribute it to the interval after the boundary.

\paragraph{\textbf{Distance counting and signed accumulation.}}
We count each input transition once by scanning the event stream in time order and attributing it to a single interval, which avoids double counting near boundaries.
Let \(N^{\mathrm{tr}}_i\) be the number of transitions attributed to interval \(i\).
Using both touch-down and touch-up events yields a spatial resolution \(\Delta d = L_{\mathrm{seq}}/2\) under an interleaved geometry in which successive transitions occur at half-pitch offsets, where \(L_{\mathrm{seq}}\) is the spatial pitch between successive reference transitions.
With the per-interval direction sign \(\sigma_i \in \{-1,0,+1\}\), we accumulate the signed displacement as
\(\hat d \;=\; \sum_{i} \sigma_i \, N^{\mathrm{tr}}_i \, \Delta d .\)

\section{Technical Evaluation}
\label{sec:technical_evaluation}
Although the time-reference mechanism tolerates irregular speed, high speeds combined with narrow electrodes (short dwell per segment) reduce per-frame capacitive change and degrade detection. 
We therefore empirically mapped accuracy across $w$ and $v$, for both configurations on two representative devices.

\subsection{Conditions}
\begin{table}[b] 
  \centering
  \small
  \caption{Experimental conditions.}
  \label{tab:exp-variables}
  \Description{A table listing the parameters for the technical evaluation, including two configurations (aligned and phase-shifted), two device types (smartphone and touchpad), four electrode widths from 1.5 to 3.0 mm, and actuation speeds ranging from 20 to 200 mm/s in 20 mm/s increments.}
  \begin{tabular}{ll}
    \toprule
    \textbf{Conditions} & \textbf{Parameters} \\
    \midrule
    Configuration & Aligned, phase-shifted \\
    Device type   & Smartphone, touchpad \\
    Width (mm)    & 1.5, 2.0, 2.5, 3.0 \\
    Speed (mm/s)  & 20–200 in 20 mm/s increments \\
    \bottomrule
  \end{tabular}
\end{table}

The experimental conditions are summarized in Table~\ref{tab:exp-variables}. 
DuoTouch works on standard capacitive touch panels. 
To identify sequence-pattern parameters that remain reliable across different form factors, we selected a smartphone (iPhone 16) as representative of medium frame rate devices (e.g., tablet PCs and smartwatches, around 60 Hz), and a touchpad (Magic Trackpad 2) as representative of higher frame rate devices (around 90 Hz). 
We evaluated both aligned and phase-shifted trace configurations. 
Figure~\ref{fig:evaluation_pattern} shows the sequence-pattern geometry.
The slider electrode width matched the electrode width in the sequence pattern \(w\); with equal widths the geometric overlap per contact is \(2w\). 
To match the off interval to \(2w\), we set the gap (edge‑to‑edge spacing) between adjacent reference electrodes to \(3w\). 
We used the 6-bit code \colorbox{gray!20}{100011} for the aligned configuration. This code contains a single on/off switch (\colorbox{gray!20}{10}) and consecutive on and off segments (\colorbox{gray!20}{00} and \colorbox{gray!20}{11}), making it representative of the patterns considered in this study. 
For the phase-shifted configuration, we used 6-bit alternation on both traces, with the input trace offset by \(w\) from the reference. 
We varied electrode width (with spacing defined relative to it) and actuation speed. We set the electrode height to 8 mm. 
Speeds ranged from 20--200 mm/s. This range covers typical finger-based interactions, and average control speeds are 100--130 mm/s~\cite{SlidingSpeed,TRing,ThumbMotorPerformance,unidirectionalSwipeGestures}.

\subsection{Setup}
\begin{figure}[h] 
  \centering
  \includegraphics[width=\linewidth]{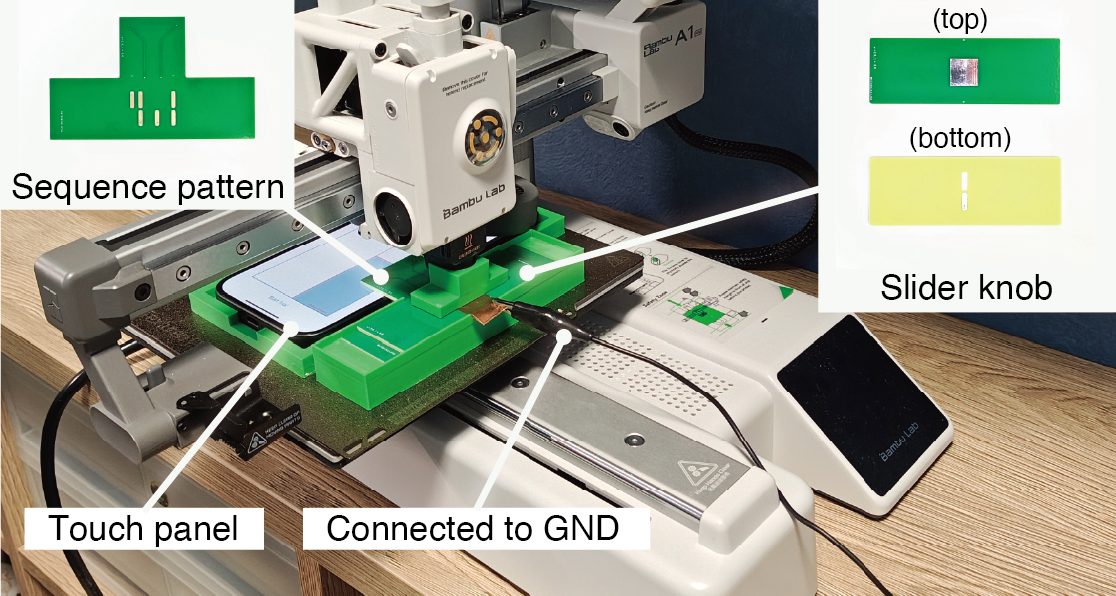}
  \caption{Technical evaluation setup: a grounded slider knob was actuated at a constant speed by the 3D printer's printhead.}
  \Description{Technical evaluation setup. A slider knob connected to ground is mounted on a 3D printer head and moves at a constant speed along the sequence pattern on a mounted device.}
  \label{fig:experimental_setup}
\end{figure}

To perform controlled interactions on a smartphone and a touchpad, we used a Bambu Lab A1 mini 3D printer (Bambu Lab). 
The printhead can be controlled via G-code at speeds from 1 mm/s to 500 mm/s in 1-mm/s increments. 
The motion controller executed constant-velocity segments at each target speed with no acceleration within the logged span. 
As shown in Fig.~\ref{fig:experimental_setup}, we mounted the smartphone and the touchpad individually on the build platform and attached a PCB with the sequence pattern to each. Guided by the footprint sizes used in prior work~\cite{Ohmic-Sticker,ShiftTouch}, we adopted square electrodes with a side length of 8 mm to ensure reliable touch detection. A slider knob with a conductive trace traveled in a straight line along the pattern; a copper-tape lead connected the slider top to ground, simulating user grounding~\cite{FindingCommonGround}. To stabilize electrical contact between the sequence pattern and the slider electrode, we removed the solder mask on the contact-facing side of the slider PCB over the electrode pad as well as the surrounding area to eliminate the mask-edge step and reduce intermittent contacts (Fig~\ref{fig:experimental_setup}). We logged touch events at the devices' default sensing rates using standard APIs~\footnote{Apple Developer (Apple) | \url{https://developer.apple.com/documentation/uikit/touches-presses-and-gestures}}: 60~Hz for the smartphone and 90~Hz for the touchpad. 
We performed 10 repetitions for every condition, yielding
2 (trace configurations) $\times$ 
2 (device types) $\times$
4 (widths) $\times$ 
10 (speeds) $\times$ 10 (trials) = 1600 total trials. 

\subsection{Results}
\begin{figure*}[t]
  \includegraphics[width=\textwidth]{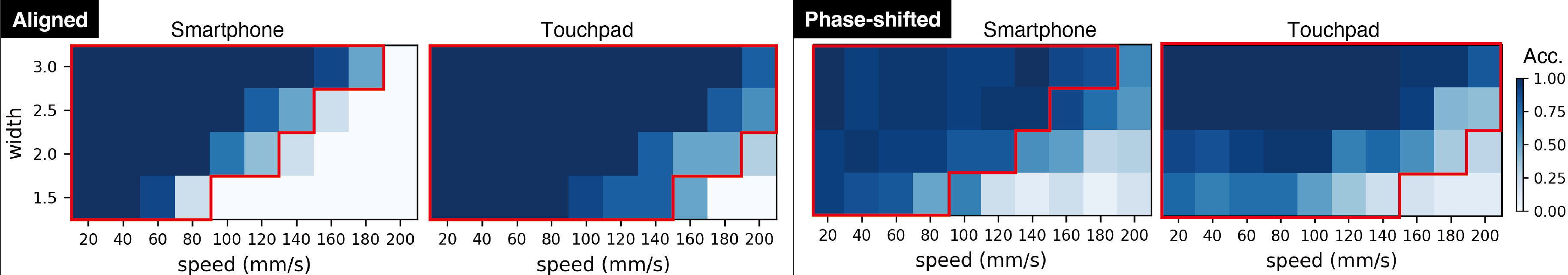}
    \caption{Recognition accuracy across speed and width for the 
    aligned and the phase-shifted configurations. 
    Red boxes indicate the sampling-limited speed bound \(v \le w f_{\mathrm{s}}\) for each frame rate. }
  \Description{Plots of recognition accuracy as a function of motion speed and trace width for aligned and phase shifted layouts. Red boxes mark the sampling limited speed bound where speed is less than or equal to width times sampling rate for each frame rate.}
  \label{fig:acc_aligned}
\end{figure*}

\paragraph{\textbf{Aligned. }}
We analyzed the collected data and only counted a trial as \emph{correct} when every bit position was classified correctly; Figure~\ref{fig:acc_aligned} (left) shows the resulting accuracy across speed and width. Red boxes indicate the sampling-limited speed bound \(v \le w f_{\mathrm{s}}\). The diagonal boundary follows this bound because each reference-bit interval must contain at least one frame. 
Within this sampling-limited bound on the smartphone (60 Hz), mean accuracy by width was 75.0\% (1.5 mm), 85.0\% (2.0 mm), 90.0\% (2.5 mm), and 93.3\% (3.0 mm).
On the touchpad (90 Hz), the corresponding means within our tested range were 92.9\% (1.5 mm), 86.7\% (2.0 mm), 94.0\% (2.5 mm), and 98.0\% (3.0 mm). 
On the 90-Hz device, the sampling-limited bound exceeds our maximum tested speed (200 mm/s) once $w \ge 2.5$ mm (225 mm/s at 2.5 mm; 270 mm/s at 3.0 mm).

\paragraph{\textbf{Phase-shifted. }}
We computed accuracy per cycle (three cycles per trial). A cycle was counted as \emph{correct} when, within its reference-bit interval, the detected order of the input touch-down and touch-up events relative to the corresponding reference transitions matched the expected direction, that is, both offsets were signed and of the correct sign. Figure~\ref{fig:acc_aligned} (right) shows the resulting per-cycle accuracy across speed and width. 
To avoid dependence on an unknown sampling phase in the worst case, reliable sign decisions require \(|\Delta t| \ge T_s\) with \(\Delta t = w/v\), which yields the same sampling-limited bound \(v \le w f_{\mathrm{s}}\) (worst-case check to avoid unknown sampling phase; this does not change the neutrality definition, which remains \(\Delta t=0\) after quantization).
Within this bound on the smartphone (60 Hz), the mean accuracy by width was 78.3\% (1.5 mm), 90.3\% (2.0 mm), 96.3\% (2.5 mm), and 94.4\% (3.0 mm). 
On the touchpad (90 Hz), the corresponding means within the tested range were 57.1\% (1.5 mm), 77.9\% (2.0 mm), 87.6\% (2.5 mm), and 97.7\% (3.0 mm).

\subsection{Discussion}
Across both devices and both decoding methods, accuracy decreased as input speed $v$ increased, with a decline for narrower electrodes $w$ and lower sampling rates $f_{\mathrm{s}}$.  
Two factors explain this trend. 
First, sampling limits temporal resolution: as $v$ rises, each reference-bit interval contains fewer sampling frames ($N \approx w f_{\mathrm{s}}/v$), so majority decisions and transition signs become unstable near the feasibility bound $v \le w f_{\mathrm{s}}$.  
Second, capacitive charging limits the usable signal. The sensing stack integrates charge over a finite per-sample window. Faster motion shortens the dwell on each electrode, and a smaller width reduces the per-frame capacitance change. 
As a result, samples near bit boundaries more often fall below the detection threshold, so the reported touch state may not change even when the geometric overlap varies. 

For the aligned configuration, the diagonal boundary in the heatmaps follows the sampling-limited condition and shifts to higher speeds on the 90-Hz device, as expected from proportionality to $f_{\mathrm{s}}$. Inside this bound, accuracy approached ceiling for the wider-width conditions, whereas the narrowest width degraded first and primarily near the boundary cells.  
Compared with aligned, the phase-shifted configuration shows lower mean accuracy and higher variance near the boundary because it requires two same-sign offsets per interval; a single neutral offset causes rejection. 
Both methods share the bound \(v \le w f_{\mathrm{s}}\) (aligned: frame scarcity; phase-shifted: offset sign instability), so device-level differences mainly reflect sampling rate.

\section{Design Guidelines and Parameter-Selection Workflow}
\label{sec:design_guidelines}
We distill the empirical and analytical results into design guidelines and apply them to the design space. 
\emph{Parameter-selection rules} (G1--G4) are grounded in 
(i) analytical results that yield the sampling-limited bound \(v \le w f_{\mathrm{s}}\) and the combinatorial constraints on assignable codes, and 
(ii) empirical accuracy maps on two devices (Fig.~\ref{fig:acc_aligned}). 
\emph{Deployment practices} (G4--G6) synthesize the implementation experience from multiple prototype runs. 

\begin{figure}[b]
\centering
\includegraphics[width=0.55\linewidth]{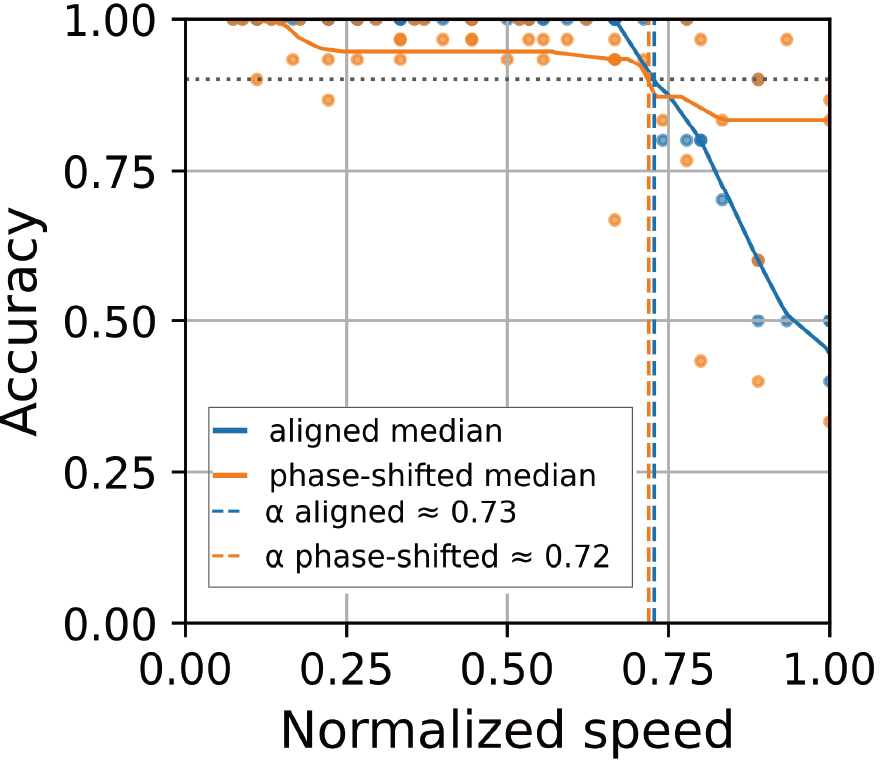}
\caption{Visual summary graph plotting accuracy and normalized speed.}
\Description{A line graph plotting decoding accuracy against normalized speed.}
\label{fig:normalized_speed}
\end{figure}

\begin{enumerate}[label=G\arabic*., leftmargin=2em]
\item \textbf{Width lower bound for the target maximum speed.}
Set $w$ such that $\frac{v_{\max}}{w f_{\mathrm{s}}} \le \alpha$.
We use $\alpha=0.73$ for the aligned configuration and $\alpha=0.72$ for the phase-shifted configuration (Fig.~\ref{fig:normalized_speed}). 
To estimate $\alpha$ from the heatmaps, for each $(w,f_{\mathrm{s}})$ we linearly interpolate along the speed axis and obtain the
normalized speed at which accuracy first crosses $0.90$ from above; 
we denote this by $s_{90}(w,f_{\mathrm{s}})$.
We aggregate $s_{90}(w,f_{\mathrm{s}})$ over widths $w\ge 2.0\,\mathrm{mm}$ and over $60$ and $90\,\mathrm{Hz}$, excluding $w=1.5\,\mathrm{mm}$ since it lies below the practical target range and consistently degrades first near the feasibility boundary, which would bias $\alpha$ downward. We take the median to produce $\alpha$ for each configuration.
\textit{Rationale:} The feasibility bound is $v \le w f_{\mathrm{s}}$; the 90\% contour lies at $s \approx 0.73$ (aligned) and $s \approx 0.72$ (phase-shifted). 
\item \textbf{Default widths for typical finger speeds.}
For $v_{\max}=100$--$130\,\mathrm{mm/s}$ (typical range of finger-based interactions~\cite{SlidingSpeed,TRing,ThumbMotorPerformance,unidirectionalSwipeGestures}), choose the width that maintains accuracy $\ge 0.90$
up to $v_{\max}=130\,\mathrm{mm/s}$. 
Accordingly, 
at $60\,\mathrm{Hz}$ use 
$w\ge 3.0\,\mathrm{mm}$ for the aligned configuration and 
$w\ge 3.0\,\mathrm{mm}$ for the phase-shifted configuration;
at $90\,
\mathrm{Hz}$ use 
$w\ge 2.0\,\mathrm{mm}$ for the aligned configuration and 
$w\ge 2.5\,\mathrm{mm}$ for the phase-shifted configuration.
\textit{Rationale:} Instantiates G1 and matches Fig.~\ref{fig:acc_aligned}. 
\item \textbf{Preference within feasible widths.}
Prefer the smaller \(w\) within the feasible region (G1/G2).
\textit{Rationale:} A smaller \(w\) increases motion resolution (phase-shifted) and reduces the minimum travel needed to complete a code (aligned). 
\item \textbf{Code length and assignment (aligned).}
For unidirectional commands, $L$ bits yield up to $2^{L}$ codes. 
For bidirectional commands, exclude palindromes and assign each reversible pair $\{p,\mathrm{rev}(p)\}$ once. 
Use the smallest $L$ that covers the required commands after exclusions. 
At runtime, issue a command only when the observed sequence exactly matches one dictionary entry. 
\textit{Rationale:} Capacity and reversibility constraints set minimal dictionary size for conflict-free assignment.
\item \textbf{One-at-a-time actuation.}
When multiple mechanisms share the same two footprints, ensure only one is mechanically actuated at a time.
\textit{Rationale:} Concurrent actuation produces overlapping sequences that violate the decoders' assumptions. 
\item \textbf{Footprint placement and occlusion.}
Position the on-screen footprints to avoid common system UI targets (e.g., back/home areas) and frequently touched app regions.\footnote{In our prototypes, the footprints were placed near the \emph{top} edge because our devices (iOS~26) position system navigation near the \emph{bottom} edge.}
\textit{Rationale:} Reduces inadvertent touches during direct pointing and preserves access to system navigation.
\end{enumerate}

\begin{figure*}[b!]
    \includegraphics[width=\textwidth]{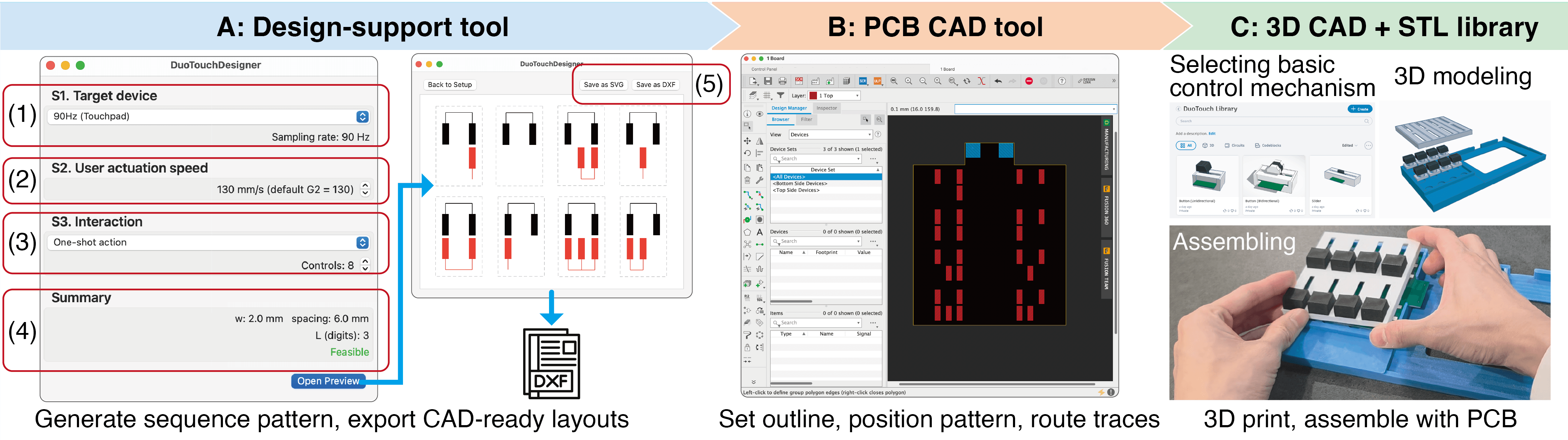}  
  \caption{Workflow for DuoTouch design and fabrication.}
  \Description{A three-stage workflow diagram: Stage A uses a design-support tool to set device parameters and export sequence patterns; Stage B involves positioning the patterns in a PCB CAD tool for board layout; and Stage C covers mechanical modeling using an STL library, 3D printing, and final assembly with the PCB.}
  \label{fig:DesignSystem}      
\end{figure*}

To select the parameter dimensions in practice, the following steps (W1--W5) operationalize G1--G4. 
\begin{enumerate}[label=W\arabic*., leftmargin=2em]
\item \textbf{Set the usage envelope.}
Select the target device (fixing $f_{\mathrm{s}}$) and the expected maximum actuation speed $v_{\max}$ for the task.
\item \textbf{Set the configuration.}
Select Aligned for discrete commands or Phase-shifted for continuous control. 
\item \textbf{Set the minimum feasible width.}
Set the smallest $w$ such that $v_{\max}/(w f_{\mathrm{s}})$ lies in the feasible region (G1/G2). 
\item \textbf{Set the intra-trace spacing.}
Set the intra-trace spacing to three times the width $w$; this balances on and off dwell times. 
\item \textbf{Set code length or travel.}
For the aligned configuration, set the bit length $L$ to the minimal value that covers the required commands after exclusions (G4). 
For the phase-shifted configuration, use the fixed pitch to set motion resolution $\Delta d = L_{\mathrm{seq}}/2$ and choose the desired physical travel as a multiple of $\Delta d$.
\end{enumerate}

\section{Design and Fabrication Workflow}
\label{sec:implementation}
In this section, we describe the DuoTouch design and fabrication workflow. The process produces PCBs that implement the sequence pattern and assembles them with 3D-printed housing and user-actuated components. 
Figure~\ref{fig:DesignSystem} shows an overview from pattern generation to PCB fabrication and final assembly.

\subsection{Design Support Tool}
To facilitate the design process, we developed a macOS application that generates, validates, and exports CAD‑ready sequence patterns.
The tool guides users through the \emph{Parameter-Selection Workflow} described in Sec.~\ref{sec:design_guidelines}, automatically setting default values to streamline configuration.
The process begins with selecting the target device class; the tool defaults the touch-sensing rate to 60 Hz for smartphones or tablets and 90 Hz for touchpads (Fig.~\ref{fig:DesignSystem}A-1). 
Next, the user sets the maximum expected actuation speed, which defaults to 130 mm/s (Fig.~\ref{fig:DesignSystem}A-2).
The user then selects an interaction type associated with a representative operation and control in Table~\ref{tab:example-mappings}. 
Based on this selection, the tool automatically assigns the Aligned configuration for discrete commands and the Phase-shifted configuration for continuous control (Fig.~\ref{fig:DesignSystem}A-3).
For Aligned inputs, the user specifies the number of controls, and the tool derives the command count to generate a minimal-length code pattern and candidate codebook.
Conversely, for Phase-shifted inputs, the tool calculates the necessary reference pitches based on the specified target travel distance.
Throughout this process, the software validates the design by computing the minimum feasible trace width and intra-trace spacing (Fig.~\ref{fig:DesignSystem}A-4).
Upon completing the configuration, users can visually inspect the generated geometry by opening the preview window; finally, the verified patterns can be exported as DXF or SVG files (Fig.~\ref{fig:DesignSystem}A-5).
The tool also generates a layout for a small PCB containing two electrodes with dimensions matching the sequence pattern, intended for use within user-actuated components. 
Users can then import these files into PCB CAD tools for final board layout and placement (Fig.~\ref{fig:DesignSystem}B).

\subsection{Fabrication and Assembly}
In a 3D CAD environment, users design the housing so that the PCB footprints, which can optionally be extended with conductive parts (e.g., a 3D-printed connector or a flexible PCB), align with the touch panel at the intended location. 
A clip-like holder presses the footprints against the touch panel and maintains capacitive coupling.
User-actuated components are integrated into the 3D-printed structure. 
Inside each component, a small two-electrode PCB is mounted at the base, and a mechanical part printed with conductive material connects the user's finger (GND) to the electrodes to establish an electrical path. 
To streamline this process, we provide a library of STL files for basic control mechanisms such as buttons, sliders, and rotary knobs, and users may also incorporate custom designs.
Finally, the housing and mechanical parts are 3D-printed and the PCB is mounted to complete the assembly (Fig.~\ref{fig:DesignSystem}C).

\begin{figure*}[b!]
  \includegraphics[width=\textwidth]{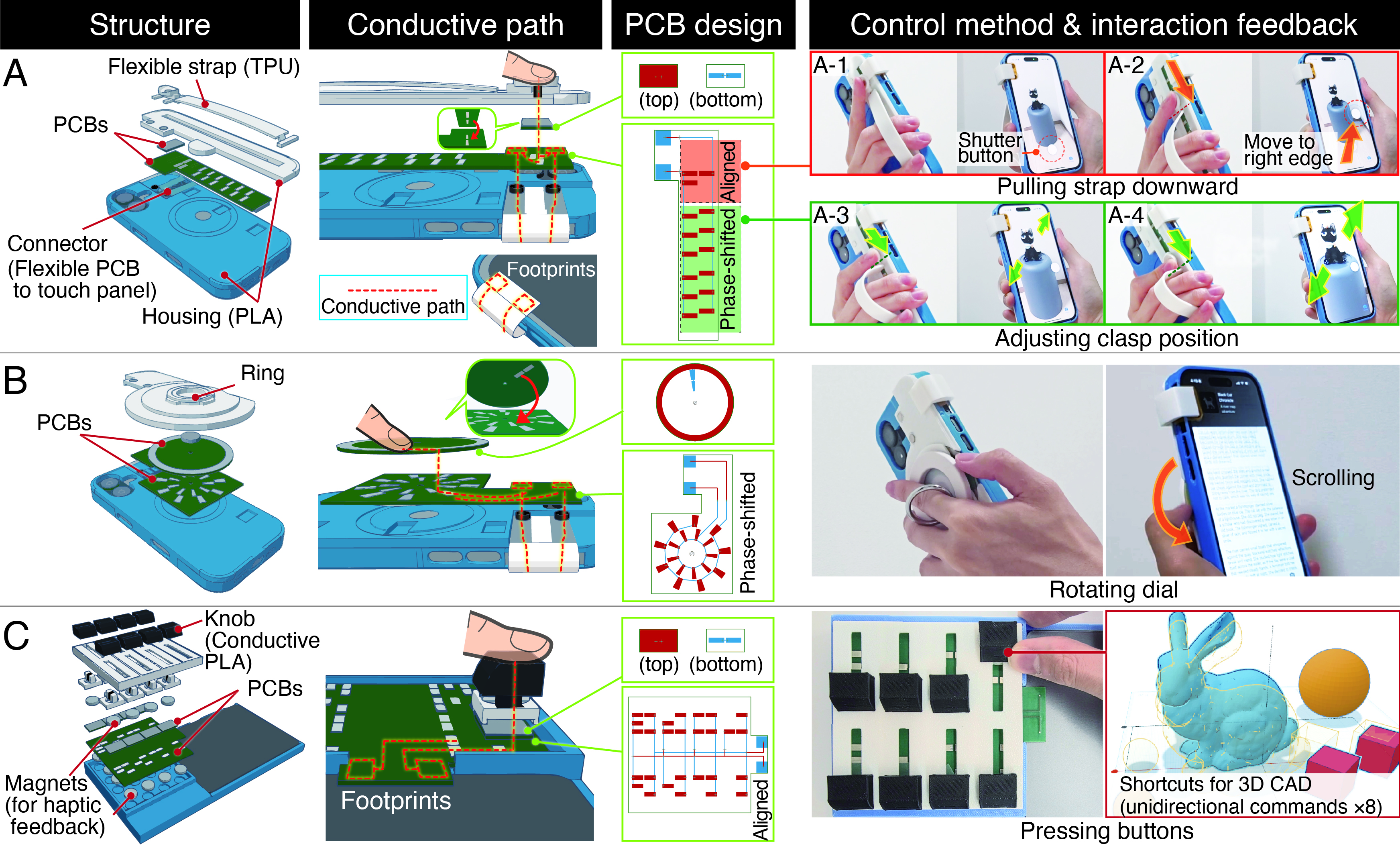}
\caption{
Applications: (A) hand strap with a slider, (B) phone ring holder with a dial, and (C) touchpad add-on with multi-button. }
\Description{Three panels show DuoTouch application prototypes. A shows a smartphone hand strap that includes a slider mechanism. B shows a phone ring holder that functions as a rotary dial. C shows a touchpad with a slim multi-button add-on; magnets inside the add-on provide a restoring force for haptic feedback.}
  \label{fig:Applications}
\end{figure*}

\section{Applications}
\label{sec:Applications}
\paragraph{\textbf{Strap-Integrated Control on a Phone Case. }}
We implemented a strap interface on the back of a smartphone case (Fig.~\ref{fig:Applications}A). The strap lies flat when not in use and extends when the user rests a finger on the clasp and pulls it downward. During the initial pull, the clasp passes over an aligned pattern and produces a binary code that activates one-handed mode. In this mode the camera shutter button moves to the right edge so that it is easy to reach with the right thumb (Fig.~\ref{fig:Applications}A-1 to A-2). 
The user can then slide the clasp up or down over a phase-shifted pattern to control camera zoom in and out (Fig.~\ref{fig:Applications}A-3 to A-4). When the strap is returned to its stored position, the clasp passes over the aligned pattern in the reverse direction and produces a different code that exits one-handed mode. 

This application combines both aligned and phase-shifted configurations, and we run both decoders concurrently. 
In our technical evaluation with a 3-mm-wide electrode on a smartphone (a setting where code recognition was reliably achieved), 96.6\% of aligned sequences showed that each transition event (both touch-down and touch-up) on the input trace coincided within the same sampling frame as its intended reference transition
\footnote{This result was obtained by analyzing the first bit (the initial \colorbox{gray!20}{1}) of the six-bit pattern (\colorbox{gray!20}{100011}).}. For certain code patterns, the aligned configuration can be confirmed within the first few bits by comparing the timing of touch-down and touch-up events between the reference and input traces. 
For example, when both traces start with \colorbox{gray!20}{11} (down direction), the alignment is confirmed from the first touch-down event. 
Alternatively, for patterns such as ref.: \colorbox{gray!20}{101} and input: \colorbox{gray!20}{010} (up direction), the alternation ensures that the events align within the first two bits, allowing early classification.

\paragraph{\textbf{Rotary Input through a Phone Ring Attachment. }}
We embedded a DuoTouch sequence pattern with a phase-shifted configuration into the base plate of a phone ring attachment, enabling rotary input on the back of a smartphone (Fig.~\ref{fig:Applications}B). As the user rotates the ring, the embedded disk slides along a circular trace, generating a continuous binary sequence. This sequence is decoded to determine motion direction and distance.
This form factor also supports one-handed use and integrates naturally with existing phone accessories. Unlike on-screen sliders or dials, the physical dial provides tactile feedback and can be operated without visual attention, making it suited for tasks such as volume adjustment, zoom control, or scrolling while holding the phone in one hand.

\paragraph{\textbf{Multi-Button Input on a Touchpad. }}
We implemented a set of eight physical buttons mounted on a touchpad (Fig.~\ref{fig:Applications}C). Pressing a button produces a unique binary sequence in the aligned configuration, which lets the system identify the pressed button. This turns the touchpad into a passive shortcut panel similar to a macropad (e.g., Stream Deck). Users can assign custom actions to individual buttons. In our prototype, we mapped 3D CAD commands (e.g., copy, paste, redo, undo, align). Because all inputs are routed through the same two footprints, the design preserves most of the touchpad surface for regular use and adds little spatial overhead.

\section{Exploratory User Study}
\label{sec:PreliminaryUserStudy}
We conducted an exploratory study with two tasks focusing on perceived experience rather than task-efficiency measures: 
Task 1, which elicited participants' initial impressions of each prototype, and Task 2, which collected everyday-use ideas for DuoTouch.
Task-based comparisons were not included here to avoid conflating the DuoTouch mechanism with early-stage UI mappings. Twelve participants took part (six female, six male; ten right-handed, two left-handed; mean age = 34.7 years, SD = 16.0). 
We explained the study purpose and procedures and obtained written informed consent in accordance with our institution's ethics guidelines.

\paragraph{\textbf{Task 1: Prototype rating. }}
Participants assessed the three applications described in Section~\ref{sec:Applications}. We refer to these prototypes as {\it strap}, {\it dial}, and {\it button}. We used a within-participant design and counterbalanced the order of the three prototypes across participants. 
Participants freely explored each prototype for approximately 5 minutes and then submitted their ratings. 
For per-prototype ratings, we utilized the User Experience Questionnaire (UEQ-S)~\cite{UEQ-S} with eight items on a 7-point scale ($-3$ to $+3$) and report \textit{Pragmatic Quality} (e.g., efficient, easy to use) and \textit{Hedonic Quality} (e.g., interesting, enjoyable) scores following the short form guidance.  

\begin{figure*}[b]
  \includegraphics[width=0.75\textwidth]{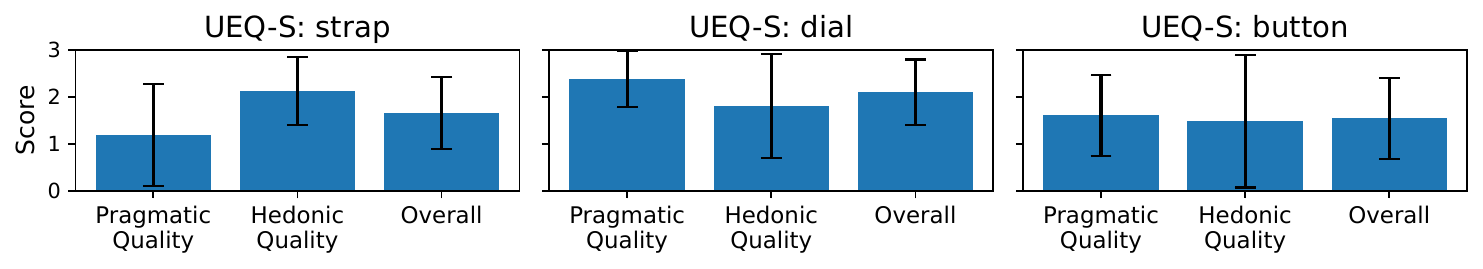}
\caption{UEQ-S mean scores for \textit{Pragmatic Quality}, \textit{Hedonic Quality}, and \textit{Overall} for each prototype; error bars show standard deviations.} 
\Description{Three side-by-side bar charts show UEQ-S mean scores for Pragmatic Quality, Hedonic Quality, and Overall for the strap, dial, and button prototypes; error bars indicate standard deviations across 12 participants.}
    \label{fig:UEQ-S}
\end{figure*}

\paragraph{\textbf{Task 2: Use case elicitation. }}
After completing Task 1, participants received a brief handout corresponding to the design space in Fig.~\ref{fig:PrimitiveInteractions}. 
The handout introduced the range of devices and input configurations that DuoTouch can support, beyond the three prototypes. 
Participants were instructed to describe three scenarios in which DuoTouch would be useful, without being limited to the three prototypes, and were encouraged to draw on the broader design space. 
Participants then completed an idea-elicitation form. For each scenario, they assigned one {\it Scenarios} label, multiple {\it Benefits} labels, and (when applicable) multiple {\it Concerns} labels.

We conducted a codebook‑based thematic analysis~\cite{AppliedThematicAnalysis} that combined deductive labels seeded by our design‑space handout with inductive labels that emerged from the data. Two authors first double‑coded a random 30\% subset of proposals. We computed Cohen's kappa on this subset at the proposal level, treating each label as present or absent. We report macro‑average $\kappa$ and the range across labels for {\it Scenarios} ($\bar{\kappa}=0.96$, min = 0.80, max = 1.00), {\it Benefits} ($\bar{\kappa}=0.89$, min = 0.71, max = 1.00), and {\it Concerns} ($\bar{\kappa}=0.97$, min = 0.73, max = 1.00). We use $\kappa$ as a descriptive index of coder agreement appropriate for a codebook workflow, not a claim of ground truth. After discussion, we reconciled differences and revised operational definitions with positive and negative examples. The two authors then independently coded the full dataset, and disagreements were resolved by consensus. Labels with fewer than two positives in the double-coded subset were excluded from $\kappa$ and are reported qualitatively.

\subsection{Results}
\paragraph{\textbf{UEQ-S.}}
Figure~\ref{fig:UEQ-S} shows the UEQ-S results.
The three prototypes achieved high scores on both \textit{Pragmatic Quality}, which 
reflects how well the system supports users in completing tasks efficiently and easily, and \textit{Hedonic Quality}, which indicates perceived stimulation and enjoyment.
According to the UEQ-S benchmark, the \textit{Overall} scores for \textit{strap} and \textit{dial} fall within the ``Excellent'' range, and that for \textit{button} falls within the ``Good'' range~\cite{UEQ-S}.

\begin{table*}[bth]
  \caption{\textit{Scenario} and \textit{benefit} labels and counts. Benefits are multi-label.}
  \Description{This table lists the scenario categories that participants proposed for DuoTouch and the benefit labels they assigned, reporting the count of scenarios or benefits in each category.}
  
  \label{tab:scenes_benefits}
  \centering
  \begin{minipage}[t]{0.5\textwidth}
    \centering
    \textbf{Scenarios (n = 36)}\\[2pt]
    \Small
    \begin{tabular}{@{}p{0.78\linewidth}r@{}}
      \toprule
      \textbf{Item} &\textbf{Count} \\
      \midrule
      Media control & 8 \\
      System UI navigation & 4 \\
      Text editing & 4 \\
      Physical interaction and sensing & 4 \\
      Sketching and annotation & 3 \\
      Camera and gallery & 2 \\
      Communication and social apps & 2 \\
      Maps and navigation & 2 \\
      \hline
      Other\textsuperscript{*} & 7\\
      \bottomrule
    \end{tabular}
    \par\smallskip
\noindent\textsuperscript{*}Other is a single coding label that comprises seven scenarios:
web browsing, game control, IoT automation, DJ control, movie editing, password input, and restaurant table ordering. These scenarios are counted under Other and are not treated as separate categories. 
  \end{minipage}\hfill
  \begin{minipage}[t]{0.48\textwidth}
    \centering
    \textbf{Benefits (n = 60)}\\[2pt]
    \Small
    \begin{tabular}{@{}p{0.78\linewidth}r@{}}
      \toprule
      \textbf{Item} & \textbf{Count} \\
      \midrule
      Ease of use & 14 \\
      Improving accuracy & 10 \\
      Expressiveness (input vocabulary and shortcuts) & 8 \\
      Speed and efficiency & 7 \\
      One-handed operation & 5 \\
      Haptic feedback & 4 \\
      Visibility (reduced screen occlusion) & 3 \\
      Accessibility & 3 \\
      Flow of operation & 2 \\
      Enjoyment & 1 \\
      Two-handed operation & 1 \\
      Eyes-free use & 1 \\
      Learnability & 1 \\
      \bottomrule
    \end{tabular}
  \end{minipage}
\end{table*}

\begin{figure*}[htb!]
  \includegraphics[width=\textwidth]{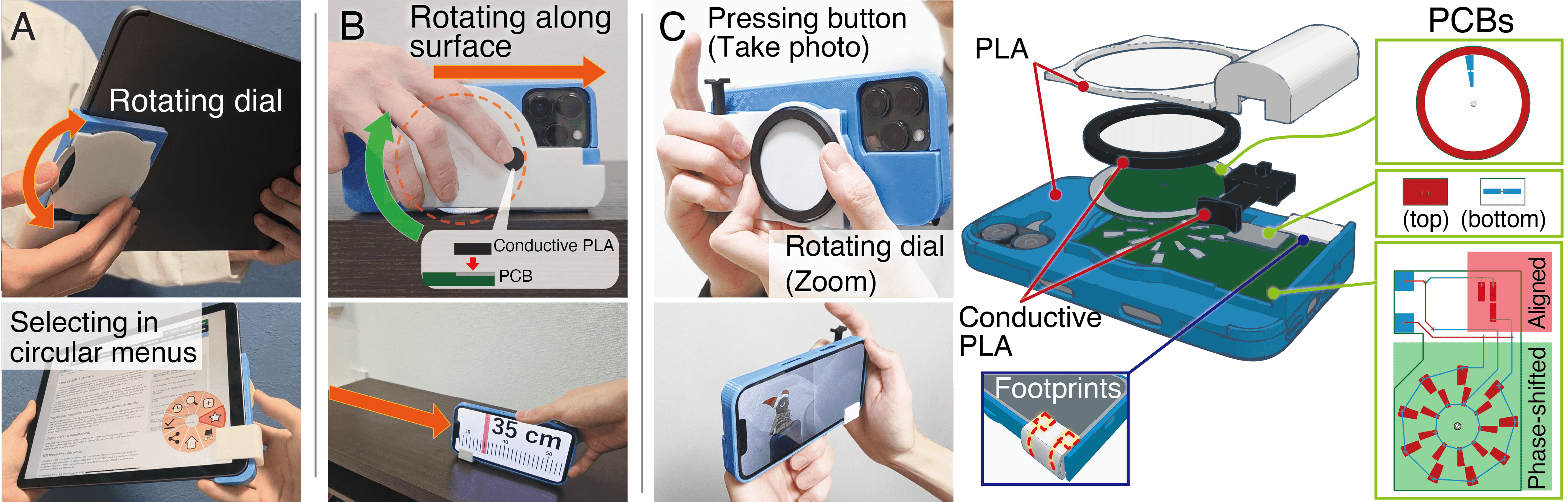}
 \caption{
   Prototypes based on participant scenarios. 
(A) Dial for discrete circular menu selection in two-handed tablet use. 
(B) Dial rolled along a surface to measure length. 
(C) Camera-shaped DuoTouch with front dial and top shutter button.
Prototypes in (A) and (B) reuse the PCB from the \emph{dial} prototype.
} 
\Description{Prototypes of DuoTouch-based devices derived from participant scenarios: a rotating dial for circular menu selection, a dial rolled along a surface to measure distance, and a camera-shaped device with a front dial for zoom and a top shutter button, reusing the same dial PCB design.}
    \label{fig:user-porposed-apps}
\end{figure*}

\paragraph{\textbf{Scenarios participants envisioned.}}
Participants proposed 36 scenarios. 
Counts for all categories are shown in Table~\ref{tab:scenes_benefits} (left).
The most frequent categories were \textit{media control} (8/36), followed by \textit{system UI navigation} (4/36), \textit{text editing} (4/36) and \textit{physical interaction and sensing} (4/36); \textit{sketching and annotation} also appeared (3/36). 
Many participants framed DuoTouch as a way to keep navigation fluid during ongoing tasks. 
Some scenarios also extended our initial definition of the design space. 
For example, although we introduced the phase-shifted configuration primarily for continuous control tasks, several participants proposed using the dial for discrete selection in circular menus (Fig.~\ref{fig:user-porposed-apps}A). 
Within the scenarios categorized as \textit{Other} (7/36), one participant envisioned using the aligned configuration of DuoTouch as a physical authentication key. While our design space focused on using the minimum code length needed for reliable command identification, this scenario instead treated long aligned sequences as password-like patterns for unlocking or authorizing actions on the host device. Prior work on tangible touch-based authentication, such as 3D-Auth~\cite{3D-Auth}, encodes authentication patterns in physical artifacts on  touch panels, and this participant scenario points to a similar application direction for DuoTouch.
Beyond app-specific actions, several scenarios targeted \textit{physical interaction and sensing} (4/36).  
Examples included rolling a dial along a surface to estimate distance in the physical world (Fig.~\ref{fig:user-porposed-apps}B) and using DuoTouch devices that mimic familiar handheld tools in everyday contexts, where the physical form and haptic feedback align with the corresponding functions on the smartphone (e.g., a camera-shaped device where rotating a circular control on the front face, in the position of a camera lens, adjusts zoom and pressing a top-side button triggers image capture (Fig.~\ref{fig:user-porposed-apps}C)).

\paragraph{\textbf{Expected benefits.}}
Counts for all categories are shown in Table~\ref{tab:scenes_benefits} (right).
\textit{Ease of use} (14/60) and \textit{improving accuracy} (10/60) were the most frequently mentioned benefits, followed by \textit{expressiveness} (8/60; input vocabulary and shortcuts) and \textit{speed and efficiency} (7/60). 
Participants reported that tasks that are usually performed by dragging on the touch panel are time-consuming and cumbersome to execute precisely with touch, and that operating them with a physical slider or dial would be easier and more accurate. Examples included using a dial to scrub in media (P8, P9, P10, P11), using a cross-key or a dial to move the text caret (P6, P10), and using a dial to adjust text selection spans (P4, P5). 
In media scenarios, which were the most frequently mentioned \textit{scenario} category, 
participants could pause or seek without bringing up the on-screen controls, which reduced the amount of GUI covering the primary content, even though DuoTouch adds small on-screen contact footprints. 
For example, P1 stated that they could pause or change position without first showing the controls, so the video was not covered by the GUI. 
Beyond supporting grip, participants also valued integrating input mechanisms into the attachment itself, noting benefits for \textit{One-handed operation} (5/60) on smartphones and \textit{Two-handed operation} (1/60) on tablets. 
For one-handed smartphone use, it was seen as affording access to controls beyond thumb reach and could enable users to access menus without swiping from out-of-reach edges (P10); 
for two-handed tablet use, it could enable operation from the attachment without shifting grip when on-screen controls were hard to reach (P5, Fig.~\ref{fig:user-porposed-apps}A). 
Participants also highlighted tactile feedback: P3 noted that the pushing or sliding is a clear indication for on or off. 
From an accessibility perspective, participants described cases of motor impairments where precisely dragging is difficult and how DuoTouch could help. P2 noted that even when precise touch input is difficult, the attachments may still be usable, for example by users with an injured finger or tremors.

\paragraph{\textbf{Concerns at adoption.}}
Counts for concerns are summarized in Table~\ref{tab:concerns}. The most frequently mentioned concerns were \textit{size} (6/29) and \textit{weight} (3/29), followed by \textit{false activation} (3/29), \textit{learning effort} (3/29), and \textit{setup effort} (3/29). Regarding \textit{size} (6/29), P6 stated that the device could become difficult to use if it were larger, and P12 stated that if the mechanism is large, it may get in the way if it is attached all the time. Regarding \textit{weight} (3/29), P2 expressed that one-handed operation is useful but would be difficult if the device were heavy. Other comments addressed \textit{false activation} (3/29) and \textit{learning effort} (3/29), stating that accidental taps might trigger unintentional functions (P1) and that the user needs to remember which function is assigned to each control (P2), respectively.

\begin{table}[h]
  \centering
  \caption{Concern labels and counts (multi-label).}
  \label{tab:concerns}
  \Description{Concern categories and counts from the study.}
  \Small
  \textbf{Concerns (n = 29)}\\[2pt]
  \begin{tabular}{@{}p{0.60\linewidth}r@{}}
    \toprule
    \textbf{Item} & \textbf{Count} \\
    \midrule
    Size & 6 \\
    Weight & 3 \\
    False activation & 3 \\
    Learning effort & 3 \\
    Setup effort & 3 \\
    \midrule
    Other & 11 \\
    \bottomrule
  \end{tabular}
\end{table}

\subsection{Discussion}
The UEQ-S results indicate a positive user experience across pragmatic and hedonic dimensions, and participants were receptive to adding input to familiar accessories (i.e., hand strap, phone ring holder). 
The \textit{button} prototype scored lower in the \textit{Overall} score (it scored Good on the UEQ-S benchmark, while the other two reached Excellent); a likely cause is learning effort. 
Several participants commented that mapping eight buttons to many shortcut commands was useful but that they forgot the mapping during use. To reduce learning effort, the hardware should provide visual and haptic cues: for example, using caps with icons or shapes that suggest common commands, keeping a consistent spatial layout, and showing labels on the screen when a control is focused or at first use. 
These cues can help participants recall the mapping without extra mental load. 
Although \textit{strap} achieved an Excellent score in the \textit{Overall} score, several participants remarked that combining the aligned and phase‑shifted modes in a single slider made operation feel complex, which may have contributed to less favorable Pragmatic impressions. This could be addressed by separating the activation and continuous‑control segments with a clear haptic boundary—for example, a stronger detent or a short magnetic stop at the activation segment—to reduce inadvertent drift.

In the idea elicitation task, participants proposed diverse scenarios, with many relating to media control, and also envisioned novel use cases that combine physical interaction and sensing. Many described drag-based tasks on touch panels as tedious and cumbersome and suggested that mapping these tasks to DuoTouch input and separating them from touch input, would improve ease, immediacy, and accuracy. 
Participants also raised concerns about size and weight, especially for \textit{strap} and \textit{dial}. These judgments were based on the assembled prototype, including the case and the ring, rather than the core mechanism alone. The DuoTouch's core mechanism (PCBs) was relatively lightweight: 7.2 g for the \textit{strap} and 17 g for the \textit{dial}. Since our study framed these prototypes for one-handed operation, this likely heightened sensitivity to mass and balance. 

Participants also raised concerns about false activation. Although our 3D-printed housings enclose the PCB traces, thereby preventing fingers from touching them during normal use, direct contact with a footprint can still trigger unintended activation. 
We also observed a failure case where a user halted a sliding button mid-pattern, resulting in an incomplete sequence. If the codebook contains prefix or substring relations, partial progress can match a shorter code and cause a false activation; when no code matches, a false negative occurs. 
This failure mode can occur even when guideline \emph{G4} is followed, since its constraint of matching exactly one dictionary entry applies only to completed sequences. 
To address this, the system could explicitly restrict code allocation to a prefix-free set, ensuring that no valid code serves as a prefix for another to prevent partial-sequence matches. Additionally, incorporating a timeout-based reset mechanism could potentially help by invalidating incomplete buffers after a certain period of inactivity. 
Along with these algorithmic improvements, providing haptic feedback at target positions could also serve as an effective physical measure to guide users toward sequence completion.

\begin{table*}[bht]
\Small
\centering
\setlength{\tabcolsep}{4pt}
\caption{Comparison between DuoTouch and finger-based touch input at the level of interaction aspects.}
\Description{This table compares DuoTouch with conventional finger-based touch input across interaction aspects including occupied screen space, command mapping, directness of control, and availability of haptic feedback.}
\label{tab:baseline-comparison}
\begin{tabular}{p{0.15\linewidth}p{0.40\linewidth}p{0.40\linewidth}}
\toprule
\textbf{Aspect} & \textbf{DuoTouch (ours)} & \textbf{Finger-based touch input} \\
\midrule
Occupied screen space 
  & Two footprints near the device edge. 
  & Icons and sliders on the display. \\
Mapping 
  & Command mappings to memorize; not visible by default. 
  & Labeled on-screen commands that are directly visible. \\
Directness of control 
  & Indirect manipulation via an attached component. 
  & Direct manipulation. \\
Haptic feedback 
  & Mechanical components can provide haptic feedback. 
  & Mostly flat glass; rich haptics require actuators in the device. \\
\bottomrule
\end{tabular}
\end{table*}

\section{General Discussion and Limitations}
\paragraph{\textbf{Baseline comparison to finger-based touch input. }}
We did not include task-based performance comparisons because early-stage, prototype-specific UI mappings and tasks would largely determine measured performance and would not generalize beyond our prototypes.
Instead, we contrast DuoTouch with conventional finger-based touch input at the level of interaction aspects (Table~\ref{tab:baseline-comparison}). 
DuoTouch keeps input in two footprints and avoids the on-screen space that finger-based shortcuts and sliders require. 
Assuming a recommended on-screen finger-touch target size of about $9.5$~mm~\cite{ANSIHFES1002007}, a toolbar of $N$ shortcuts occupies roughly $N \times 9.5$~mm. 
DuoTouch relies on user-memorized command mappings, whereas finger touch uses labeled on-screen commands that are easier to discover. 
As an attached component, DuoTouch also provides indirect control that can be operated when directly reaching the content is difficult, such as during one-handed use or on larger devices. 
Direct finger input acts at the locus of visual feedback and can be preferable for spatial tasks such as drawing. 
DuoTouch components can incorporate mechanical buttons and sliders that provide haptic feedback, whereas the glass surface of a touch panel provides limited tactile structure. 
Taken together, these trade-offs suggest that DuoTouch and conventional finger-based touch input are complementary rather than competing. 
Table~\ref{tab:baseline-comparison} provides a starting point for designers to peruse the practical advantages and costs of direct finger input, DuoTouch, and combinations of both for a given task and device. 
In future work, this aspect-level comparison could be complemented by controlled task-based studies once more mature application mappings are available.

\paragraph{\textbf{Materials and fabrication.}}
To minimize screen occlusion, transparent conductors (e.g., ITO film) are a promising option for footprints. 
This improves visibility but does not remove coupling effects; any conductive footprint can perturb the touch-sensing stack and shift reported touch coordinates. Material choice also determines fabrication and assembly. 
Our prototypes use PCBs assembled into 3D-printed housings, which provides repeatability and geometry control but requires manual assembly and narrows the material choice. 
The DuoTouch sensing principle is material-agnostic: the sequence pattern can also be formed with conductive films or with conductive filament co-printed with PLA housings, enabling single-process fabrication. 
Conductive filament has higher resistivity and rougher contact than PCB traces, which can lower per-sample amplitude at high speed and shrink the feasible region set by the sampling bound \(v \le w f_{\mathrm{s}}\). 
We estimate a material margin \(\alpha_{\mathrm{mat}}\) by sweeping a short straight pattern over the target speed range, then choose the width using \(v_{\max} \le \alpha_{\mathrm{mat}}\, w\, f_{\mathrm{s}}\). 
Our design-support tool currently emits verified sequence patterns; extending it to co-fabrication that emits both mechanical and conductive parts is future work. 

\paragraph{\textbf{Adoption trade-offs. }}
While DuoTouch requires a more involved fabrication process (i.e., 3D printing and PCB manufacturing) than rapidly fabricated interfaces constructed from materials like paper or PET films, this approach targets scenarios where mechanical robustness is a priority. Unlike transient interfaces prone to wear through repeated interaction, DuoTouch's structural integrity ensures that mechanical responses remain stable over extended use. Furthermore, compared to active electronic alternatives such as Bluetooth controllers, DuoTouch's passive nature eliminates the need for batteries and pairing, lowering long-term maintenance costs. 
Beyond the demonstrated prototypes, our encoding strategy provides a formalized approach for creating robust physical controllers tailored to individual application needs without the complexity of active electronics. 
This framework, supported by our analytical sampling bounds, defines a middle ground between rapid but fragile low-fidelity prototypes and fixed commercial peripherals, offering the mechanical reliability of a dedicated device with the flexibility of bespoke hardware. 

\begin{table}[b]
  \centering
  \caption{Unit prices by finish at two quantities. $Q$ denotes the number of boards (single design). Shipping and taxes excluded.}
  \label{tab:finish-cost}
  \Description{This table reports indicative unit prices for DuoTouch PCBs at two order quantities and three surface finishes, showing how manufacturing cost varies with the chosen finish.}
  \Small
  \begin{tabular}{lcc}
    \toprule
    \textbf{Finish} & \textbf{Unit price at $Q{=}25$} & \textbf{Unit price at $Q{=}50$} \\
    \midrule
    Lead-free HASL & \$1.71 & \$1.23 \\
    ENIG           & \$2.70 & \$1.73 \\
    ENEPIG         & \$10.28 & \$5.48 \\
    \bottomrule
  \end{tabular}
\end{table}

\paragraph{\textbf{Durability and cost.}}
Our prototype PCBs use lead-free HASL as the surface finish. It is inexpensive but offers lower abrasion resistance than ENIG or ENEPIG. For longer service life, ENIG or ENEPIG can be considered at higher cost. To inform practical choices, Table~\ref{tab:finish-cost} reports indicative quotes we obtained from PCBWay~\footnote{\url{https://www.pcbway.com/}} (August~2025) for our board under fixed conditions~\footnote{60$\times$100\,mm, FR-4, 1.6-mm thickness, 1\,oz Cu, min trace/space 6/6\,mil (approx. 0.15 mm), min drill 0.2\,mm, green solder mask, white silkscreen, no edge connector.} while varying only the quantity (single design) and the surface finish. These results show the cost–durability trade-off rather than universal prices. 
Motivated by this trade-off, we performed an automated repeated sliding test using the same setup as in the technical evaluation: one lead-free HASL board was slid against a counterpart for 1000 cycles at a speed of 200\,mm/s. After 1000 repetitions, all electrodes remained functional, and the resistance from the footprint to each electrode segment averaged below 0.05\,$\Omega$. These preliminary results suggest that lead-free HASL can maintain sufficient conductivity for our application, though broader wear studies with multiple specimens, environmental factors, and comparisons to ENIG and ENEPIG should be done in the future.

\paragraph{\textbf{Simultaneous use and grounding.}}
DuoTouch shares the same limitations as other touch panel attachments that extend touch interactions (e.g., ~\cite{BackTrack,KirigamiKeyboard,ShiftTouch}). It does not support simultaneous input from multiple components. 
In addition, DuoTouch relies on the user's body as ground to generate touch input. As a result, touch detection may fail in cases such as when the user's fingers are dry. 
This can potentially be mitigated by using device settings such as glove mode to lower the detection threshold, or by connecting DuoTouch's electrodes to the device chassis ground instead of relying on the user's body, as demonstrated in previous work~\cite{Ohmic-Sticker,PaperTouch}. 
Finally, DuoTouch requires a small movement before a decision can be made because recognition relies on transitions rather than contact. 
While this prevents classification at the instant of first touch, it aligns with a primary benefit of touch-panel attachments: providing physical feedback and guidance during input.

\paragraph{\textbf{High interaction speeds.}}
DuoTouch supports tangible controls such as sliders and dials and was stable at typical finger speeds in our evaluation. 
Some uses demand higher speeds, such as rapid fader flicks or continuous jog wheel spins (one participant proposed a DJ controller scenario during the idea elicitation task in Sec.~\ref{sec:PreliminaryUserStudy}). 
At such rates, limited per-sample integration and the panel sampling rate could cause misses near the sampling-limited bound \(v \le w f_{\mathrm{s}}\). 
This bound is an analytical guideline in this work rather than a value reported in prior studies, yet the same temporal dependence applies to interfaces that rely on sampled touch events. 
For example, ExtensionSticker~\cite{ExtensionSticker} achieves continuous scrolling by propagating contact across strip array footprints, so continuity at high speeds is in principle limited by sampling fidelity. 
Higher touch-sensing rates enlarge the feasible region roughly proportional to \(f_{\mathrm{s}}\), so the limit is a device constraint rather than a method constraint.

\section{Conclusion}
We presented DuoTouch, a passive attachment that adds tangible input to capacitive touch panels using two contact footprints. 
On a smartphone and a touchpad we validated decoding across trace width and motion speed and derived a sampling-limited bound that links these factors to sampling rate.
We mapped these results to design guidelines and a design-support tool that outputs fabrication-ready sequence patterns. We embedded DuoTouch in a hand strap, a phone ring holder, and macropad add-ons, and an exploratory user study indicated benefits in ease of use, accuracy, expressiveness, haptic feedback, and accessibility, with size and weight as main concerns. 
These results establish the feasibility of DuoTouch and guide deployment of tangible input on touch panels.

\begin{acks}
We would like to thank Buntarou Shizuki and the anonymous reviewers for their insightful discussions and helpful feedback on this work. 
\end{acks}

\bibliographystyle{ACM-Reference-Format}
\bibliography{sample-base}
\end{document}